\documentstyle[12pt]{article}

\begin{document}


\def\a#1#2#3#4{{\,{#1}_{#3}{#2}_{#3}\cdots\,{#1}_{#4}{#2}_{#4}}}

\def\b #1{\mbox{\footnotesize\boldmath$#1$\normalsize}}

\def\sobd #1#2{\stackrel{\mbox{\boldmath$\rightarrow$}}{#1#2}}

\def\sobi #1#2{\stackrel{\mbox{\boldmath$\leftarrow$}}{#1#2}}

\def\sobr #1#2{\overline{#1#2}}

\def\subr #1#2{\underline{#1#2}}

\def\SG{\mbox{\Large$\sigma$\normalsize}}

\def\DG{\mbox{\large$\delta$\normalsize}}

\def\ind#1#2#3{^{\phantom{#1} #2}_{#1\phantom{#2} #3}}

\def\an#1#2{\a{#1}{#2}{1}n}

\def\ra#1{{#1\llap{/}}}

\def\vk#1#2#3#4#5{\underline{\vc{#1}{#2}{#3}}\vc{#1}{#4}{#5}}

\def\vc#1#2#3{{\,{#1}_{#2}\cdots\,{#1}_{#3}}}

\title{The Extended Loop Representation of \\
Quantum Gravity}

\author{ \\ \\ \small
{\bf Cayetano Di Bartolo} \\
\small Departamento de F\'\i sica, Universidad Sim\'on Bol\'\i var, \\
\small Apartado 89000, Caracas 1080-A, Venezuela. \\ \\
\small {\bf Rodolfo Gambini and Jorge Griego }\\
\small Instituto de F\'{\i}sica, Facultad de Ciencias,\\
\small Trist\'an Narvaja 1674, Montevideo, Uruguay.}

\date{June 16, 1994}
\maketitle
\vspace{0.5cm}
\abstract{A new representation of Quantum Gravity is developed. This
formulation is based on an extension of the group of loops. The enlarged
group, that we call the Extended Loop Group, behaves locally as an
infinite dimensional Lie group. Quantum Gravity can be realized on the
state space of extended loop dependent wavefunctions. The extended
representation generalizes the loop representation and contains this
representation as a particular case. The resulting diffeomorphism and
hamiltonian constraints take a very simple form and allow to apply
functional methods and simplify the loop calculus. In particular we show
that the constraints are linear in the momenta.  The nondegenerate
solutions known in the loop representation are also solutions of the
constraints in the new  representation. The practical calculation
advantages allows to find a new solution to the Wheeler-DeWitt equation.
Moreover, the extended representation puts in a precise framework some
of the regularization problems of the loop representation. We show that
the solutions are generalized knot invariants, smooth in the extended
variables, and any framing is unnecessary.}

\newpage
\section{Introduction}
The formulation of General Relativity in  terms  of  the  Ashtekar
variables has opened new perspectives in the canonical quantization
program of gravity \cite{A1,A2}. The new set of canonical variables
introduced by Ashtekar are the  triads $E^{ax}_i$  (the projections of
the tetrads onto  a three surface)  and a complex SU(2) connection
$A^i_{ax}$. The fundamental result  of  this  approach  is   that  the
constraint equations emerging from the hamiltonian formulation of
General Relativity become  polynomial functions  of the variables.
Besides, the formalism (that uses a connection as the configuration
variable)  casts General  Relativity in  a fashion that closely
resembles Yang-Mills theories. This fact allows  to import several
useful techniques from Yang-Mills theories  into General Relativity. One
of particular importance is the loop representation \cite{GT,RS,RG}.

The loop representation provides a geometric description of the
hamiltonian formulation of the theory (gauge theories and quantum
gravity) in terms of loops. The loop representation can be constructed
by means of the noncanonical algebra of a complete set of gauge
invariant operators that act on a state space of loop wavefunctions
$\psi(\gamma)$ \cite{GT,RS}. Once the complete set of invariant
operators is realized on the space of loops, the action of any other
gauge invariant operator (like the hamiltonian) can be obtained from
them. Another equivalent way to obtain the loop representation is
through the loop transform. The loop transform connects the states
between the connection and the loop representation and choosing a factor
ordering, the quantum operators acting in the connection state space can
be translated to the loop wavefunctions. This procedure explicitly shows
the role of holonomies as the basic building blocks of any loop
dependent object.

The introduction of the loop representation for quantum gravity allows
to immediately code the invariance under spatial diffeomorphisms in the
requirement of knot invariance. Also for the first time a large class of
solutions to the Wheeler-DeWitt equation has been found in terms of
nonintersecting knot invariants \cite{RS}. It is not clear however how
to make these solutions correspond to nondegenerate metrics \cite{BP91}
(a possible solution is the idea of ``weaves" \cite{ARS}). Another
alternative is to consider knot invariants of intersecting loops and
solve the Wheeler-DeWitt equation in loop space \cite{RG}. The
definition of many of these states is complicated by regularization
ambiguities. Since loops are one dimensional objects living in a three
dimensional manifold, they naturally lead to the appearance of
distributional expressions. In particular the few knot invariants for
which we have analytic expressions require the introduction of
regularizations (framings) in the case of intersecting knots. Some
invariants even require a regularization for smooth loops
\cite{W89,GMM}.

There are good reasons to consider an  extension  of  the loop
representation. From  a   mathematical   point of view, the group of
loops is not a Lie group  and  it   is   not even known how to give  a
manifold   structure   to   the   loop space. An extension of the loop
space with  the  structure  of  an infinite dimensional Lie group has
been  recently  proposed \cite{GGG}. The usual  group of loops is a
subgroup of this  extended  group and generalized holonomies may
be defined in the  extended  space. Moreover, since the extended
representation uses fields instead of distributional objects,
the regularization difficulties associated with the wavefunctions in the
loop representation disappear. On the other side, Marolf  \cite{Maro}
has recently studied  the  equivalence between the connection and the
loop representation in the case   of (2+1)-gravity. More precisely, he
has shown that some  problems and ambiguities arise in the kernel of the
loop transform. For instance, in the  boost sector the loop transform
has  a very  large  non-trivial  kernel. He also noticed that the
introduction of extended loops  allow  to remove many  of  these
problems. Even though the introduction of the extended representation is
not mandatory in this case (the equivalence may be recovered in the loop
representation by introducing a nontrivial measure in the loop
transform \cite{AL}), extended loops seem to give a very simple and
natural solution to this problem.

In this article we develop a formalism that allows to represent quantum
General Relativity in the extended space. The general ideas of the
extended representation have been discussed in reference \cite{PRL}. This
representation can be viewed as a generalization of the loop
representation. Due to the particular extended group we use, the
formalism can be considered as the most simple and general of all
possibles for Quantum Gravity and have the characteristic to present the
constraint equations in a very simple form and to increase their power
of resolution. Moreover, the regularization problem associated with the
formal action of the constraint operators simplifies considerably in
this approach, being the most relevant result the removal of the typical
ambiguities associated with the framing dependence that the knot
invariants have in the loop representation.

We organize the article as follows: in section 2, we make a brief review
of the definition and properties of the loop coordinates, that preceded
and motivated the definition of the extended group. In section 3, the
extended loop group is introduced in a formal way. We only do here a
quick review of the loop coordinates and the extended loop group in
order to make the article moderately selfcontained. A more complete
treatment  of these subjects  can  be  found  in reference \cite{GGG}.
Section 4 is dedicated to the construction of the extended loop
representation of Quantum Gravity. In first place we analyze the general
properties of the wavefunctions and the operators in the extended
representation. We show that the constraints are linear in
the momenta. In section 4.1 the diffeomorphism constraint is formulated
in the extended space. In section 4.2 we consider the realization of the
hamiltonian constraint. The reduction of the extended hamiltonian
constraint to the corresponding in the loop representation is performed in
section 4.3. This procedure allows to clarify the meaning of some of the
new ingredients that appear in the extended hamiltonian constraint. In
section 5 we develop the formal calculus considering the action of the
hamiltonian on the second coefficient of the Alexander-Conway
generalized knot polynomial. We will verify that this state is
annihilated by the extended hamiltonian constraint. Also a new solution
of the hamiltonian constraints is reported. The issue of the
regularization is considered in section 6. In section 6.1 we show that
the wavefunctions are smooth functionals of the extended variables (in a
restricted diffeomorphism invariant domain). The regulated
diffeomorphism and hamiltonian constraints are analyzed in section 6.2.
Some conclusions and final comments are included in section 7.

\section{The Loop Coordinates}
Due to their simple behavior under gauge transformations, holonomies
have been widely used in the description of gauge theories. Holonomies
can be viewed  as an homomorphism going from a group structure defined in
terms of equivalence classes of closed curves onto a Lie group G. Each
equivalence class is called a loop and the group structure defined by them
is called the group of loops. The group of loops is the basic underlying
structure to all the nonlocal formulations of gauge theories in terms of
holonomies \cite{GT}.

As it was just mentioned, among these formulations we find the loop
representation, based on a quantum representation of the
hamiltonian gauge theory in terms of loops. In the loop representation
wavefunctions are functionals of loops and they are connected with
the states in the connection representation  by the loop transform
\begin{equation}
\psi(\gamma)=\int d_{\mu}[A] \psi(A) W_A (\gamma)
\end{equation}
where
\begin{equation}
W_A (\gamma) =  Tr[ H_A (\gamma)] = Tr[ P
\exp{(\oint_{\gamma} dy^{a} A_{ay})}]
\end{equation}
is the Wilson loop functional. All the gauge invariant information
present in the theory can be retrieved from the holonomy. This means that
the only information we really need to know from loops is the
one used in the definition of $H_A (\gamma)$. One can write the holonomy in
the following way
\begin{equation}
H_A (\gamma) = 1 +  \sum_{n=1}^{\infty} \int dx^3_1
\ldots  dx^3_n  A_{a_1x_1} \ldots A_{a_nx_n} X^{a_1 x_1 \ldots a_n x_n}
(\gamma)
\end{equation}
where the loop dependent objects $X$ are given by:
\begin{eqnarray}
&& X^{a_1 x_1 \ldots a_n x_n}(\gamma) =  \nonumber \\ &&
\oint_\gamma dy_n^{a_n} \! \ldots \! \oint_\gamma dy_{1}^{a_1} \delta
(x_n\!-y_n) \! \ldots \! \delta(x_1\!-y_1) \Theta_\gamma(o,y_1,\! \ldots
\! ,y_n)
\end{eqnarray}
The $\Theta$ function orders the points along
the contour starting at the origin of the loop $o$. These relationships define
the  $X$  objects  of  ``rank"  n, that we   call    the multitangents of
the  loop $\gamma$.   They  behave  as multivector  densities under
general  coordinate transformations.  The fundamental property is that no
more information from the loop  is  needed  in  order  to compute  the
holonomy than what is present in the multitangents fields of all rank.

It is convenient to introduce the following notation for the multitangent
fields
\begin{equation}
X^{\b \mu}(\gamma) = X^{\mu_1 \ldots \mu_n}(\gamma)
= X^{a_1\,x_1\ldots a_n\,x_n}(\gamma)
\end{equation}
where $\b \mu$ indicates the set of indexes
$(\mu_1,\ldots, \mu_n)$ and $\mu_i$ represents the pair of variables
$(a_i,x_i)$, with $a_i = 1, 2, 3$ and $x_i \in {\cal{R}}^3$.

The X's are not independent quantities, they  obey  two  kinds  of
constraints: the algebraic and differential constraints.

The algebraic constraints arise from the properties of the $\Theta$
function under the interchange of the order of the indexes and have the
following general form
\begin{equation}
X^{\vk \mu 1k{k+1}n} \equiv
\sum_{P_k} X^{P_k(\mu_1 \ldots \mu_n)} = X^{\mu_1\ldots\mu_k} \,
X^{\mu_{k+1}\ldots \mu_n}
\end{equation}
where the  sum  goes over  all  the  permutations of  the  $\mu$
variables which preserve the ordering of  $(\mu_1,\ldots,\mu_k)$ and
the $(\mu_{k+1},\ldots,\mu_n)$  among  themselves.

The differential constraints ensure  that  $H_A (\gamma)$  has  the
correct transformation properties under gauge transformations and
can be derived directly from the definition of the multitangent
fields as line integrals of distributions along closed curves.  They
are given by
\begin{equation}
\partial_{\mu_i} X^{\mu_1 \ldots \mu_i \ldots \mu_n} =
\left[ \delta(x_i-x_{i-1}) - \delta(x_i-x_{i+1}) \right]
 X^{ \mu_1 \ldots \mu_{i-1} \; \mu_{i+1} \ldots \mu_n }
\label{diffconst}
\end{equation}
where $\partial_{\mu_i} \equiv \partial / \partial x_i^{a_i}$.
Notice that the differential constraint carries information about the
origin of the loop because for $i=1$ or $i=n$, the points $x_0$ and
$x_{n+1}$ has to be taken as the basepoint $o$ of the loop.

The first idea was to find a set of independent quantities that
completely specify the loop dependent information contained in the
holonomy. The solution of the constraints can be outlined in the
following way:
\begin{eqnarray*}
\mbox{\bf X fields}
\longleftrightarrow & \stackrel{\mbox{\footnotesize{solving
AC}}}{\mbox{\footnotesize{keeping DC}}}& \longleftrightarrow \mbox{\bf F
fields} \longleftrightarrow \mbox{\footnotesize{solving DC}}
\longleftrightarrow \mbox{\bf Y fields} \nonumber \\ X^{\b
\mu}(\gamma)\hspace{-1cm} &=&\hspace{-1cm} [\, exp\, F(\gamma)\, ]^{\b
\mu}\, , \hspace{0.5cm} F^{\b \mu}(\gamma) =
{\mbox{\Large{$\sigma$}}}^{\b \mu}{_{\b \nu}} \, Y^{\b \nu}(\gamma)
\nonumber \\ F^{\b   \mu}(\gamma)\hspace{-1cm}   &=&\hspace{-1cm}
[\,ln\,  X  (\gamma)\,]^{\b  \mu}\, ,  \hspace{0.8cm}  Y^{\b
\mu}(\gamma)   = \delta_T^{\b \mu}{_{\b \nu}} \, F^{\b \nu}(\gamma)
\end{eqnarray*}
The $F$'s are multivector density fields that satisfy
the homogeneous algebraic constraint and the differential constraint.
The $Y$'s satisfy both homogeneous algebraic and differential
constraints and define the loop coordinates associated with the loop
$\gamma$. $\delta_T$ is the projector on  the  space  of transverse
functions. The matrix ${\mbox{\Large{$\sigma$}}}$ is
nondiagonal and  generates nontrivial representations of the
diffeomorphism group. Their explicit expressions can be found in
reference \cite{GGG}. In the above equations a  generalized Einstein
convention  of sum  over  repeated indexes is assumed, given by
\begin{eqnarray} &&
A_{\b \mu}\, B^{\b \mu} = \sum^{\infty}_{n=0} \, A_{\mu_1 \ldots \mu_n}
\, B^{ \mu_1 \ldots \mu_n}  \\ && A_{\mu_i}\, B^{ \mu_i} =
\sum_{a_{i}=1}^{3} \int d^3 x_i A_{a_i}(x_i) \,B^{a_i}(x_i)
\end{eqnarray}

The definition of the exponential and the inverse operation that
connects the multitangent fields with the algebraic free coordinates
involves a particular composition law between the components
of the fields. Explicitly,
\begin{equation}
[\, exp\,F(\gamma)\, ]^{\b \mu} \; \stackrel{def}{=}
\; \sum_{k=0}^{\infty} \frac{1}{k!} [ F \times
\stackrel{k-times}{\cdots} \times F]^{\b \mu}
\end{equation}
with
\begin{equation}
[F \times F]^{\, \mu_1 \ldots \mu_n } \;
\stackrel{def}{=} \;
\sum_{i=1}^{n-1} F^{\, \mu_1 \ldots \mu_i }\, F^{\, \mu_{i+1}
\ldots \mu_n }
\end{equation}
This composition law is associative and has the important property
that satisfies the differential constraint if all $F$'s do.

The loop coordinates have several interesting applications at  the level
of the loop representation of  gauge  theories  and  quantum gravity.
What is more important, they allow  to  show  that  there exist an
infinite dimensional manifold  with  a  local  Lie  group structure
associated with the  loop  space.  We  will now  proceed to introduce
this group.

\section{The Extended Loop Group}
Consider a set of arbitrary multitensor densities of any rank
and construct with them the following  vector-like object ${\bf E}$
\begin{equation}
{\bf E} \, =   (\, E, \, E^{\, \mu_1}, \, \ldots , \,
E^{\,\mu_1\ldots \mu_n},\, \ldots ) \; \equiv (\, E, \, \vec{E})
\end{equation}
where $E$ is a real number and $E^{\, \mu_1 \ldots  \mu_n}$   (for
any $n  \neq 0$) is an arbitrary multivector  density  not restricted by
any constraint. The set of all ${\bf E}$'s has the structure of a
vector   space, denoted as ${\cal E}$.

We can introduce a product law in ${\cal E}$  in the following way:
given two vectors ${\bf E}_1$ and ${\bf E}_2$, we define ${\bf E}_1
\times {\bf E}_2$ as the vector with components
\begin{equation}
{\bf E}_1 \times {\bf E}_2 \; = \; (\,E_1 E_2, \, E_1 \vec{E_2}  +
 \vec{E_1}E_2  +  \vec{E_1} \times \vec{E_2})
\end{equation}
where  $\vec{E_1}  \times  \vec{E_2}$   is   given   by,
\begin{equation}
(\vec{E}_1  \times  \vec{E}_2)^{\,\mu_1\ldots \mu_n}   =
\sum^{n-1}_{i=1}
E_1^{\,\mu_1\ldots \mu_i}E_2^{\,\mu_{i+1}\ldots \mu_n}\;\;.
\end{equation}
We see that the $\times$-product is an extension of  the  composition
law between the algebraic free coordinates introduced before that includes
the zero rank  component  of  the  vector.  In  fact,  it  can  be
written as
\begin{equation}
({\bf E}_1 \times {\bf E}_2)^{\,\mu_1\ldots \mu_n}   =
\sum^{n}_{i=0} E_1^{\,\mu_1\ldots \mu_i}E_2^{\,\mu_{i+1}\ldots
\mu_n} \label{gpro}
\end{equation}
with the convention
\begin{equation}
E^{\,\mu_1\ldots \mu_0}   =   E^{\,\mu_{n+1}\ldots \mu_n} = E
\end{equation}
The product law is associative and distributive  with  respect  to
the addition of vectors. It has a null element (the  null  vector)
and a identity element, given by
\begin{equation}
{\bf I} = (\, 1,\, 0,\, \ldots ,\, 0, \, \ldots)
\end{equation}
An inverse element exists for all vectors with nonvanishing zero rank
component. It is given by
\begin{equation}
{\bf E}^{-1} \; = \; E^{-1} {\bf I} \; + \;
\sum^{\infty}_{i=1}
(-1)^i E^{-i-1}
({\bf E} \, - \, E {\bf I})^i \label{inverse}
\end{equation}
such that
\begin{equation}
{\bf E} \times {\bf E}^{-1} \; = \; {\bf E}^{-1} \times {\bf E}
 \; = \; {\bf I} \;\; .
\end{equation}

The set of all vectors with  nonvanishing zero rank component
forms a group with the $\times$-product law.

Consider now a subset $\cal{X} \subset \cal{E}$ whose elements ${\bf X}
\equiv (\,X,\,\vec{X})$ obey some supplementary conditions related to
the differential and algebraic constraints. Introducing different types
of conditions one can define different subgroups.
We consider here the following three basepointed subgroups of the
extended group, denoted by ${\cal D}_{o}$, ${\cal  M}_{o}$ and ${\cal
X}_{o}$. They are defined in the following way
\begin{eqnarray*}
{\cal    D}_{o} &  \rightarrow     &  [\vec{X}]
\mbox{  satisfies    the
differential constraint and $X \neq 0$} \\
{\cal M}_{o}  &  \rightarrow     &      [\vec{X}]
\mbox{ satisfies
the differential constraint and} \\
& & \hspace{0.6cm} \mbox{
$[X^{-1}]^{\mu_1\ldots   \mu_n}  =    (-1)^{n}X^{\mu_n\ldots \mu_1}$
and  $X =1 $} \\
{\cal X}_{o} & \rightarrow &  [\vec{X}]
\mbox{ satisfies     the  differential
and  algebraic constraints} \\
&&\hspace{0.6cm} \mbox{ and $X=1$ }
\end{eqnarray*}
For each of  these  sets one can check that they are closed under  the
group composition law. It is straightforward to see that the algebraic
constraint also implies the condition satisfied by the elements of  ${\cal
M}_o$. These groups satisfy then the following inclusion
relation
\begin{equation}
{\cal X}_{o} \subset {\cal M}_{o} \subset {\cal D}_{o}
\end{equation}
Furthermore, the group of loops is a subgroup of the  group
${\cal X}_{o}$, since any multitangent field ${\bf X}(\gamma)$
is contained in  ${\cal X}_{o}$ and they obey the following
relationship
\begin{equation}
{\bf X}(\gamma_{1} \circ \gamma_{2})= {\bf
X}(\gamma_{1})\times
{\bf X}(\gamma_{2})
\end{equation}
where the circle  indicates  the  group  product  between loops.

An important property  of  the   constraints  is  that  {\em  any}
mutitensor density $X^{\b \mu}$  that  satisfies them can be used to
generate  a  {\em    gauge covariant}     quantity
\begin{equation}
H_A({\bf X})= A_{\b \mu} X^{\b \mu}
\end{equation}
When restricted to the multitangents $X(\gamma)$  associated  with
loops, the resulting object is the holonomy. It is  this  property
that allows to extend loops to a more general structure.   One
can in general deal with  arbitrary  multitensor   densities   $X$
(not necessarily related to loops)  and   construct  with  them
gauge invariant objects.  The multitensor  densities   need    not
to be   distributional   functions as   the multitangents  associated
with a  loop.  They  could  be ordinary functions  on  the
manifold.

Matrix representations of the above groups can be  generated
through   the   generalized   holonomies   associated    with    a
general connection $A_{ax}$. They satisfy
\begin{equation}
H_A({\bf X_1})  H_A({\bf X_1}) = H_A({\bf X_1} \times {\bf X_2})
\end{equation}
The differential constraint imposed on $\bf X$ assures   $H_A({\bf
X})$ to be a gauge covariant quantity. The trace of  the  extended
holonomy, that defines a generalized Wilson  functional,     is  a
gauge  invariant  quantity    for  any     ${\bf  X}\in      {\cal
D}_{o}$. Since  one  can  represent  any  gauge  invariant  object
using the $X$'s, one  can  represent  the theory  {\em   entirely}
in   terms of  $X$'s.

An   extended   representation   may    be  introduced   for   any
subgroup of the larger constrained group ${\cal D}_{o}$. In each case we
will obtain a gauge invariant representation of a theory with particular
properties derived from the constraints. For Quantum Gravity (where the
connections are elements of the SU(2) algebra) the matrices $H_A ({\bf
X})$ would belong to the general group $GL(2,c)$ when ${\bf X} \in
{\cal D}_{o}$, whereas for the other subgroups the generalized
holonomies would be elements of the SU(2) group (in fact, the condition
satisfied by the elements of ${\cal M}_o$ is the weakest for this
property to hold). We shall  consider here the simplest case of a
representation with   wavefunctions defined on the larger  group ${\cal
D}_{o}$. However  it   is important to notice that    more restricted
representations could be relevant and it is still unclear which of  them
behaves as the dual of the connection representation.

\section{The Extended Loop Representation of \protect \\
Quantum Gravity}
Let us start by considering some general properties of the wavefunctions
in the extended representation. These wavefunctions are related  to the
states  in  the  connection representation  through   the
generalized loop transform
\begin{equation}
\psi({\bf X})=\int d_{\mu}[A] \, \psi(A) \, W_A ({\bf X})
\end{equation}
with
\begin{equation}
W_A ({\bf X}) =  Tr[ A_{\b \mu}] \, X^{\b \mu}
\end{equation}
The generalized Wilson functional satisfies a set of identities that
correspond to the Maldestam identities for the SU(2) algebra. The
cyclic property of the traces implies
\begin{equation}
W_A ({\bf X}_1 \times {\bf X}_2 )  =
W_A ({\bf X}_2 \times {\bf X}_1 )
\end{equation}
while taking into account the specific properties of the SU(2) algebra, one
gets
\begin{equation}
W_A ({\bf X})  =  W_A (\overline{{\bf X}})
\label{W}
\end{equation}
and
\begin{equation}
W_A ({\bf X}_1 ) \, W_A ({\bf X}_2 ) = W_A ({\bf  X}_1
\times {\bf X}_2 ) +
W_A ({\bf X}_1 \times \overline{{\bf X}}_2 )
\end{equation}

\vspace{0.3cm}
\noindent
where
\begin{equation}
\overline{X}^{ \mu_1 \ldots \mu_n} \equiv (-1)^n
X^{ \mu_n \ldots \mu_1}
\end{equation}
The identities satisfied by the Wilson functional are carried over
the wavefunctions in a direct way. We get
\begin{eqnarray}
 \psi ({\bf X}_1 \! \times \! {\bf X}_2 ) &=& \psi({\bf  X}_2  \!
\times \! {\bf X}_1 ) \\
 \psi({\bf X}) &=& \psi(\overline{{\bf X}}) \label{psi} \\
 \psi({\bf X}_1 \! \times \! {\bf X}_2 \! \times
\!  {\bf  X}_3 ) +
\psi({\bf X}_1 \!\!\!\!\! & \times &\!\!\!\!\! {\bf X}_2 \! \times \!
\overline{{\bf X}}_3  )  =  \nonumber \\
&& \hspace{-1cm} \psi({\bf X}_2 \! \times \!
{\bf X}_1 \!\! \times \! {\bf X}_3 ) +
\psi({\bf X}_2 \! \times \!
{\bf X}_1 \!\! \times \! \overline{{\bf X}}_3 )
\end{eqnarray}
Notice that due to the cyclic property of the traces any reference
to the basepoint $o$ is lost in $\psi$. Furthermore, equations (\ref{W})
and (\ref{psi}) imply that $W_A$ and $\psi$ are functions
of the combination
\begin{equation}
R^{\mu_1 \ldots \mu_n} = \frac{1}{2} \left[  X^{\mu_1
\ldots  \mu_n}  + (-1)^n X^{\mu_n \ldots \mu_1} \right]
\end{equation}
where the ${\bf R}$'s satisfy the following symmetry
property under the inversion of the indexes
\begin{equation}
R^{\mu_1 \ldots  \mu_n}  = (-1)^n R^{\mu_n \ldots \mu_1}
\label{simr}
\end{equation}

The linearity of the extended holonomy in the multivector components
$X^{\mu_1 \ldots \mu_n}$ induces the same property on the wavefunctions.
This means that $\psi{(\bf X})$ takes the general form
\begin{equation}
\psi({\bf   X})   =   D_{\b \mu}   \,    X^{\b \mu}
\end{equation}
The  coefficients  $D_{\mu_1  \ldots  \mu_n}$  contain   all   the
information about $\psi({\bf  X})$  and  satisfy
a set of constraints  that  may  be  derived  from  the  Maldestam
identities. They are
\begin{eqnarray}
&& \hspace{2.2cm} D_{\mu_1 \ldots \mu_n} =
D_{(\mu_1 \ldots \mu_n)_{c}}  \\
&&\hspace{2.2cm}D_{\mu_1\ldots \mu_n} = (-1)^n D_{\mu_n \ldots \mu_1} \\
&& D_{\mu_1\ldots \mu_k \mu_{k+1} \ldots \mu_n} + (-1)^k D_{\mu_k\ldots \mu_1
\mu_{k+1} \ldots \mu_n} = \nonumber \\
&& \hspace{1cm} k^{-1} D_{(\mu_1\ldots \mu_k)_c \, \mu_{k+1}
\ldots \mu_n} + (-1)^k k^{-1} D_{(\mu_k\ldots \mu_1)_c \, \mu_{k+1}
\ldots \mu_n} \;\, {\mbox{\footnotesize{for all {\it k}}}}
\end{eqnarray}
where $c$ indicates the cyclic combination of indexes. The linearity is
a remarkable property of the wavefunctions in the extended
representation. Notice that all the wavefunctions known in the loop
representation for quantum gravity have this property when they are
written in terms of the multitangents fields. Moreover, this property
will be inherited also by the operators that we can construct in the
extended representation. In general, the linearity over the
wavefunctions could be imposed by means of the ``linearity constraint"
${\cal L}$
\begin{equation}
{\cal L} ({\bf X}') \, \psi({\bf X}) \equiv X'^{\,\b \mu \b \nu}
\frac{\delta^2}{\delta X^{\b \mu} \delta X^{\b \nu}} \, \psi({\bf
X}) = 0
\end{equation}
where ${\bf X}'$ is any object that satisfies the differential
constraints. Any observable of the theory has to commute with the
linearity constraint. This means that the action of any observable
reduces to evaluate the wavefunction on a new element.

In the quantum version of the canonical formulation of General Relativity,
the classical variables are promoted to operators that act on
wavefunctionals of the connection in the following way
\begin{eqnarray}
{\hat A}^i_{ax} \psi (A) &=& A^i_{ax} \psi (A) \\
{\hat E}^{ax}_i \psi (A) &=& \frac{\delta}{\delta A^i_{ax}} \psi (A)
\end{eqnarray}
The diffeomorphism and hamiltonian constraint in this
representation are
\begin{eqnarray}
{\tilde{\cal  C}}_{ax}  &=& \frac{\delta}{\delta A^i_{bx}}F^{\,i}_{ba}  (x)
\\
{\tilde{\cal  H}} (x)  &=&  \epsilon^{ijk}
\frac{\delta}{\delta A^k_{bx}}\frac{\delta}{\delta A^j_{ax}}F^{\,i}_{ba}  (x)
\end{eqnarray}
Notice  that we choose the factor ordering that  puts  the  triads to
the  left. It is from this factor  ordering   that   one   can obtain
the  loop representation via the loop transform in a simple way. One
can see that with this factor  ordering  the  regularized diffeomorphism
constraint generates  infinitesimal  diffeomorphism transformations  on
the wavefunctions \cite{BGPNP}.

We now analyze how to go from the connection to the extended
representation by means of the loop transform. In spite that the
extended loop transform is only formally defined, one can use it as an
heuristic method to generate the constraints. In particular, the algebra
of a set of operators in the connection representation can be correctly
implemented in the linear space of extended wavefunctions using the
formal transform. The same thing happens in the loop representation.

\subsection{The diffeomorphism constraint}
Let  us  consider first the  diffeomorphism  constraint operator. The
action of this constraint on the wavefunctions $\psi({\bf R})$ is
defined by the following expression
\begin{equation}
{\cal  C}_{ax} \psi({\bf R}) =
\int d_{\mu}[A] \,W_A  ({\bf  R})\,[\,{\tilde{\cal   C}}_{ax}
\psi(A)\,]
\end{equation}
The  constraint acting on $\psi(A)$ can be  applied  on   the
generalized  Wilson  functional integrating (formally) by  parts. So
\begin{equation}
{\cal  C}_{ax} \psi({\bf R})
= \int d_{\mu}[A] \, \psi(A)  \,
 [  F^{\,i}_{ab}  (x)  \frac{\delta}{\delta  A^i_{bx}}  W_A
({\bf R}) ]
\label{diff}
\end{equation}
In order to calculate the square bracket in the last equation
we introduce some suitable notation. Let $\DG^{\b \alpha}_{\b \beta}$ be
defined as
\begin{equation}
\DG^{\b \alpha}_{\b \beta} = \left\{
\begin{array}{ll}
\delta^{\alpha_1}_{\beta_1} \cdots \delta^{\alpha_n}_{\beta_n}
\;, & \mbox{ if $n({\b \alpha}) = n({\b \beta}) = n \geq 1$} \rule{0mm}{6mm}
\\ 1 \;, &\mbox{ if $n({\b \alpha}) = n({\b \beta}) = 0$} \rule{0mm}{6mm}
\\ \rule{0mm}{6mm}
0\;, &\mbox{ in other case}
\end{array}
\right. \label{deltag}
\end{equation}
where $n({\b \alpha})$ is the number of indexes of the set ${\b
\alpha}$. The functional derivative of any product of $A$'s can be
written as
\begin{equation}
\frac{\delta}{\delta A^i_{bx}} (A_{\b \alpha}) =
A_{\b \mu} \tau^i A_{\b \nu} \,\DG^{\b \mu \, bx \, \b \nu}_{\b \alpha}
\end{equation}
where the $\tau$'s are the generators of the SU(2) algebra. Taking the
trace in the above expression we get
\begin{equation}
\frac{\delta}{\delta A^i_{bx}} Tr[A_{\b \alpha}] =
Tr[\tau^i \, A_{\b \beta} ] \, \DG^{\b \beta}_{\b \nu \b \mu}
\,\DG^{\b \mu \, bx \, \b \nu}_{\b \alpha} =
Tr[\tau^i \, A_{\b \beta} ] \, \DG^{(bx\, \b \beta)_c}_{\b \alpha}
\label{dertra}
\end{equation}
The curvature tensor can be written in the following way
\begin{equation}
F_{ab} (x) = A_{\b \nu} {\cal F}_{ab}{^{\b \nu}}(x)
\label{rcursiva}
\end{equation}
where ${\cal  F}_{ab}$  represents  the  following element  of  the
algebra of the group
\begin{equation}
{\cal  F}_{ab}{^{\b \nu}}(x)  =  \delta_{1, n(\b \nu)} \,{\cal
F}_{ab}{^{\nu_1}}  (x)  + \delta_{2, n(\b \nu)}\,
{\cal F}_{ab}{^{\nu_1 \nu_2}} (x)
\end{equation}
with
\begin{eqnarray}
&& {\cal F}_{ab}{^{a_1 x_1}}(x)   =
\delta_{a\,b}^{a_1 \!\;  d} \; \partial_d
\, \delta(x_1 - x) \\
&&  {\cal  F}_{ab}{^{a_1  x_1   \!,   \,   a_2   x_2}}   (x)     =
\delta_{a\,b}^{a_1\, a_2}
\; \delta(x_1 - x) \, \delta(x_2 - x)
\end{eqnarray}
Using (\ref{dertra}) and (\ref{rcursiva}) we obtain the following
expression for the action of the diffeomorphism constraint on the
generalized Wilson functional
\begin{eqnarray}
F^{\,i}_{ab}(x) \frac{\delta}{\delta A^i_{bx}} Tr[A_{\b \alpha}] \, R^{\b
\alpha}
&=& Tr[F_{ab}(x) \,A_{\b \beta}] \,\DG^{(ax\, \b \beta)_c}_{\b
\alpha}\,R^{\b \alpha}
\nonumber \\
&=&
Tr[A_{\b \rho}] \, \DG^{\b \rho}_{\b \nu \b \beta} \,
{\cal  F}_{ab}(x){^{\b \nu}} \,R^{(bx \, \b \beta)_c}
\label{difwil}
\end{eqnarray}
The $\DG$-matrix allows to write the group product defined in
(\ref{gpro}) in the following way
\begin{equation}
({\bf E}_1 \times {\bf E}_2)^{\b \rho}   =
\DG^{\b \rho}_{\b \nu \b \beta} \, E_1^{\b \nu}\, E_2^{\b \beta}
\label{gpro2}
\end{equation}
Notice that in particular
\begin{equation}
({\bf \DG}_{\b \nu} \times {\bf \DG}_{\b \beta})^{\b \rho}   =
\DG^{\b \rho}_{\b \nu \b \beta}
\end{equation}
where ${\bf \DG}_{\b \alpha}$ is the ``vector" with components $({\bf
\DG}_{\b \alpha})^{\b \mu} = \DG^{\b \mu}_{\b \alpha}$. Introducing then
(\ref{difwil}) in (\ref{diff}) and using (\ref{gpro2}) we obtain
\begin{eqnarray}
{\cal  C}_{ax} \psi({\bf R})
 &=& \int d_{\mu}[A] \psi(A) \,
 Tr[A_{\b \rho}] \left[ {\cal  F}_{ab}(x) \times {\bf R}
^{(bx)} \right]^{ \b \rho} \nonumber \\
&=& \psi({\cal  F}_{ab}(x) \times {\bf R}^{(bx)} )
\label{extdif}
\end{eqnarray}
where
\begin{equation}
[{\bf R}^{(bx)}]^{\b \mu} = R^{(bx) \b \mu}
\equiv R^{(bx \, \b \mu)_c}
\end{equation}
The diffeomorphism constraint reduces to evaluate the wavefunction on a
new object given by the group product between an element of the algebra
and a cyclic combination of elements of the group. This combination
satisfies the differential constraint with respect to the ${\b \mu}$
indexes basepointed at $x$.

\subsection{The hamiltonian constraint}
A similar procedure can be followed to built up the hamiltonian
constraint in the extended representation. In this case we have to use
the properties of the SU(2) algebra in order to take into account the
two derivatives that appear in ${\tilde{\cal H}} (x)$. We have now
\begin{equation}
{\cal H} (x) \,
\psi({\bf R}) = \int d_{\mu}[A] \, \psi(A)  \, \epsilon^{ijk} [
F^{\,i}_{ba}  (x)  \frac{\delta}{\delta  A^j_{bx}} \frac{\delta}{\delta
A^k_{ax}}  W_A ({\bf R}) ]
\label{hamcon}
\end{equation}
{}From (\ref{dertra}) we get the following expression for the second
functional derivative
\begin{eqnarray}
&&\frac{\delta}{\delta  A^j_{bx}}
\frac{\delta}{\delta A^k_{ax}} Tr[A_{\b \alpha}]  = Tr[\tau^k
\frac{\delta}{\delta  A^j_{bx}} A_{\b \beta}] \,
\DG^{(ax\, \b \beta)_c}_{\b \alpha} = \nonumber \\
&& Tr[\tau^k A_{\b \mu} \tau^j A_{\b \nu}]\,
\DG^{\b \mu \,bx\,\b \nu}_{\b \beta} \,\DG^{(ax\, \b \beta)_c}_{\b \alpha}=
Tr[\tau^k A_{\b \mu} \tau^j A_{\b \nu} ]
\,\DG^{(ax\, \b \mu\,bx\,\b \nu)_c}_{\b \alpha}
\end{eqnarray}
In order to include this result in (\ref{hamcon}) we need the following
well known property of the SU(2) matrices
\begin{equation}
\epsilon^{ijk} Tr[\tau^k A_{\b \mu} \tau^j A_{\b \nu} ] =
Tr[\tau^i A_{\b \nu} ]\,Tr[A_{\b \mu} ] -
Tr[A_{\b \nu} ] \,Tr[\tau^i A_{\b \mu} ]
\end{equation}
The product between traces of SU(2)
matrices can be merged in a combination of traces in the following way
\begin{equation}
Tr[A_{\b \mu} ] \, Tr[A_{\b \nu}] = Tr[A_{\b \mu} \,
A_{\b \nu}] + (-1)^{n({\b \nu})} Tr[A_{\b \mu} \, A_{{\b \nu}^{-1}}]
\end{equation}
where if ${\b \nu} = (\nu_1, \ldots , \nu_n)$, then ${\b \nu}^{-1} =
(\nu_n, \ldots , \nu_1)$. This fact enables to write the above result
in the form
\begin{equation}
\epsilon^{ijk} Tr[\tau^k A_{\b \mu} \tau^j A_{\b \nu} ] =
(-1)^{n({\b \mu})} Tr[\tau^i A_{\b \nu} \,A_{{\b \mu}^{-1}} ] -
(-1)^{n({\b \nu})} Tr[\tau^iA_{{\b \nu}^{-1}}\,A_{\b \mu}]
\end{equation}
We then have
\begin{eqnarray}
&&\epsilon^{ijk}
F^{\,i}_{ba}  (x)  \frac{\delta}{\delta  A^j_{bx}} \frac{\delta}{\delta
A^k_{ax}}\, Tr[A_{\b \alpha}]= \nonumber\\
&&(-1)^{n({\b \mu})}\,
Tr[F_{ba}(x) A_{\b \nu \b \mu}] \left\{
\DG^{(ax\, \b \mu^{-1} \,bx\,\b \nu)_c}_{\b \alpha}
- (-1)^{n({\b \mu}+{\b \nu})}\,
\DG^{(ax\, \b \mu \,bx\,\b \nu^{-1})_c}_{\b \alpha} \right\} =\nonumber \\
&&(-1)^{n({\b \mu})}\,
Tr[F_{ba}(x) A_{\b \nu \b \mu}] \left\{
\DG^{(bx\, \b \nu \,ax\,\b \mu^{-1})_c}_{\b \alpha}
+ (-1)^{n({\b \mu}+{\b \nu})}\,
\DG^{(\b \mu \,ax\,\b \nu^{-1}\,bx)_c}_{\b \alpha} \right\} =\nonumber \\
&&(-1)^{n({\b \mu})}\,
Tr[A_{\b \beta \b \nu \b \mu}]\, {\cal  F}_{ab}{^{\b \beta}}(x)\,
\DG^{(ax\, \b \nu \,bx\,\b \mu^{-1})_c}_{\b \gamma} \left\{
\DG^{\b \gamma}_{\b \alpha}
+ (-1)^{n({\b \gamma})}\,
\DG^{\b \gamma^{-1}}_{\b \alpha} \right\}
\end{eqnarray}
So
\begin{eqnarray}
&&\hspace{-1.1cm}\epsilon^{ijk}
F^{\,i}_{ba}  (x)  \frac{\delta}{\delta  A^j_{bx}} \frac{\delta}{\delta
A^k_{ax}}\,W_A ({\bf R}) = \nonumber \\
&& \hspace{1cm} 2\, (-1)^{n({\b \mu})}\,
Tr[A_{\b \beta \b \nu \b \mu}]\, {\cal  F}_{ab}{^{\b \beta}}(x)\,
\DG^{(ax\, \b \nu \,bx\,\b \mu^{-1})_c}_{\b \gamma}\,R^{\b \gamma} =
\nonumber \\
&&\hspace{1cm} 2\,(-1)^{n({\b \mu})}\,
Tr[A_{\b \alpha}]\, \DG^{\b \alpha}_{\b \beta \b \rho}\,
{\cal  F}_{ab}{^{\b \beta}}(x)\,\left[ \DG^{\b \rho}_{\b \nu \b \mu}\,
R^{(ax\, \b \nu \,bx\,\b \mu^{-1})_c}\right]
\label{hamwil}
\end{eqnarray}
where in the first step we have used the symmetry property
(\ref{simr}) of the $R$'s under the inversion of the indexes. The
expression between square brackets defines a specific combination
of $R$'s, that we denote
\begin{equation}
[{\bf   R}^{(ax, \,bx)}]^{\b \rho} =
R^{ (ax, \,bx) \b \rho} \equiv
({\bf \DG}_{\b \nu} \times
{\bf \DG}_{\b \mu})^{\b \rho}\,(-1)^{n({\b \mu})}\,
R^{(ax\, \b \nu \,bx\,\b \mu^{-1})_c}
\end{equation}
Explicitly
\begin{equation}
R^{(ax, \,bx) \rho_1 \ldots \rho_n} = \sum_{k=0}^{n}
(-1)^{n-k}\,
R^{(\, ax \, \rho_1 \ldots \rho_k \, bx \, \rho_n \ldots
\rho_{k+1})_c}     \label{rsobre}
\end{equation}
An important fact is that this combination satisfies the differential
constraint with respect to the ${\b \rho}$ indexes basepointed at $x$.
It satisfies besides the following property
\begin{equation}
R^{(ax, \,bx) \b \rho^{-1}} =
(-1)^{n({\b \rho})}\,R^{(bx, \,ax) \b \rho}
\end{equation}
Equation (\ref{hamwil}) can then be written
\begin{eqnarray}
\epsilon^{ijk}
F^{\,i}_{ba}  (x)  \frac{\delta}{\delta  A^j_{bx}} \frac{\delta}{\delta
A^k_{ax}}\,W_A ({\bf R})\!\! &=&\!\!
2\,Tr[A_{\b \alpha}]\,
({\bf \DG}_{\b \beta} \times
{\bf \DG}_{\b \rho})^{\b \alpha}\,
{\cal  F}_{ab}{^{\b \beta}}(x)\,R^{(ax, \,bx) \b \rho}
\nonumber \\
&=& \!\!2\,Tr[A_{\b \alpha}]\,
({\cal  F}_{ab} \times {\bf R}^{(ax, \,bx)})^{\b \alpha}
\end{eqnarray}
and from it we conclude
\begin{equation}
{\cal H} (x) \, \psi({\bf R}) =
2 \,\psi ({\cal F}_{ab} (x) \times {\bf   R}^{(ax, \,bx)})
\label{ham}
\end{equation}
Also in this case the action of the hamiltonian constraint reduces
to evaluate the wavefunction on a new element. As it was just
mentioned, this is a general property of the operators in the
extended representation due to the linearity of the wavefunctions.
In fact, the last expression can be written in terms of the
functional derivative with respect to the ${\bf R}$ variables
\begin{equation}
{\cal H} (x) \, \psi({\bf R}) =
2 \,\left[ {\cal F}_{ab} (x) \times {\bf   R}^{(ax, \,bx)}
\right]^{\b \mu} \,
\frac{\delta}{\delta R^{\b \mu}} \, \psi ( {\bf   R})
\end{equation}
Notice that in order for this expression to be well defined is necessary that
the term contracted with the functional derivative satisfy the differential
constraint, as it happens in this case. This result explicitly shows that
the hamiltonian is linear in the ``momenta" $P_{\b \mu}\equiv \delta
/\delta {\bf R}^{\b \mu}$.

The new object where the wavefunction is evaluated involves a
combination of multitvector density fields with two indexes fixed at the
point where the hamiltonian is acting and the other indexes having an
specific alternating order. We will show in the next section that this
alternating order of the indexes is related to the reroutings of a loop
when the above expression is particularized to loops. The appearance of
a rerouting is typical of the loop representation and plays a crucial
role in the quantum gravity case \cite{reruteo}.

\subsection{From the extended to the loop
representation}
As it was shown in Sect. 3, the group  of loops ${\cal L}_o$ is a
subgroup of the extended group ${\cal D}_o$. The extended version of the
constraints (\ref{extdif}) and (\ref{ham}) can be particularized to
${\cal L}_o$ simply by substituting ${\bf R} \rightarrow {\bf
R}(\gamma)$. We analyze here in detail the case of the hamiltonian
constraint.

It is a well known fact that the hamiltonian constraint in the loop
representation has only a nontrivial action on intersecting
loops \cite{RS,RG,BP91}. We suppose then that at the point $x$ the loop
$\gamma$ intersects itself $p$ times; that is to say, $\gamma$ has
``multiplicity" $p$ at $x$. We start with some suitable notation to take
into account this fact.

If the loop $\gamma$ has multiplicity $p$ at $x$ one can write it in
the following way
\begin{equation}
\gamma_{xx} = \gamma^{(1)}_{xx} \circ  \gamma^{(2)}_{xx} \circ
\cdots \circ \gamma^{(p)}_{xx}
\end{equation}
We denote by $[\gamma_{xx}]_i^{i+j}$ the following composition of loops
basepointed at $x$
\begin{equation}
[\gamma_{xx}]_i^{i+j} = \gamma^{(i)}_{xx} \circ
\cdots \circ \gamma^{(i+j)}_{xx}
\end{equation}
Let us suppose that the loop named $\gamma^{(1)}_{xx}$ contains the
origin $o$ of the loops. Then
\begin{equation}
\gamma_o = \gamma^{(1)}{_o^x} \circ  [\gamma_{xx}]_2^p  \circ
\gamma^{(1)}{_x^o}
\end{equation}
Here, $\gamma^{(1)}{_o^x}$ represents the portion of $\gamma^{(1)}$ from
the origin $o$ to the point $x$. The loop $\gamma_o$ is completely
described by the multitangent fields $X^{\b \mu}(\gamma_o)$ of all ranks.
As we know, these fields satisfy both algebraic and differential
constraints. Besides, these objects have another property related to the
possibility of writing a loop as the composition of open paths. In general
\begin{equation}
X^{\mu_1 \ldots \mu_n}(\gamma_o) = \int_{\gamma_o} dz^{a_i} \delta(x_i -
z) X^{\mu_1 \ldots \mu_{i-1}}(\gamma_o^z)\,X^{\mu_{i+1}
\ldots \mu_n}(\gamma_z^o)
\end{equation}
Suppose now that the index $\mu_i$ is fixed at the point $x$. Then
\begin{eqnarray}
&& X^{\mu_1 \ldots \mu_i \, ax \,\mu_{i+1} \ldots \mu_n}(\gamma_o) =
\nonumber \\
&& \sum_{m=1}^{p}
X^{\mu_1 \ldots \mu_i}(\gamma^{(1)}{_o^x} \circ  [\gamma_{xx}]_2^m) \,
T_m^{ax} \,
X^{\mu_{i+1} \ldots \mu_n}([\gamma_{xx}]_{m+1}^p \circ \gamma^{(1)}{_x^o})
\end{eqnarray}
where $T_m^{ax}$ is the tangent at $x$ when the loop passes the time $m$
to this point and the following convention is assumed:
$[\gamma_{xx}]_{m+1}^m \approx {\em i}_{xx}$, with ${\em i}_{xx}$ the
null path. The above expression can be easily generalized to the case of
any number of indexes fixed at $x$.

In order to evaluate ${\bf R}^{(ax, \,bx)}(\gamma_o)$ we have
to use the explicit expression of this object in terms of the
multitangents fields. We have
\begin{eqnarray}
&&R^{(ax, \,bx) \mu_1 \ldots \mu_n} =
{\textstyle{1 \over 2}} \sum_{k=0}^{n}
\sum_{l=0}^{k} (-1)^{n-k} [ X^{\mu_{l+1} \ldots \mu_k \, bx
\,\mu_n \ldots \mu_{k+1} \, ax \, \mu_1 \ldots \mu_l}
\nonumber \\
&& \hspace{5cm} + (-1)^n X^{\mu_l \ldots \mu_1 \, ax
\,\mu_{k+1} \ldots \mu_n \, bx \,\mu_k \ldots \mu_{l+1}} ]
\nonumber \\ &&  + \sum_{k=0}^{n}  \sum_{l=k}^{n} (-1)^{n-k} [
X^{\mu_{l} \ldots \mu_{k+1} \, ax \,\mu_1 \ldots \mu_k \, bx
\, \mu_n \ldots \mu_{l+1}} \nonumber \\
&& \hspace{5cm} + (-1)^n
X^{\mu_{l+1} \ldots \mu_n \, bx \,\mu_k \ldots \mu_1 \, ax
\,\mu_{k+1} \ldots \mu_l} ] \label{res}
\end{eqnarray}
One can write the above expression in a more compact and useful form
introducing the following combinations of $X$'s:
\begin{equation}
X^{(ax,  \sobd bx )  \b   \mu} \,
\equiv \, \sum_{k=0}^{n} (-1)^{n-k} X^{(\, ax \, \mu_1
\ldots \mu_k \, bx \, \mu_n \ldots \mu_{k+1})_c} \label{vec1}
\end{equation}
and
\begin{equation}
X^{(ax, \sobi bx ) \b \mu}
\equiv \, \sum_{k=0}^{n} (-1)^k X^{(\, ax \, \mu_k
\ldots  \mu_1  \,  bx  \,   \mu_{k+1} \ldots \mu_n)_c}
\label{vec2}
\end{equation}
These objects satisfy definite symmetry properties under the
inversion of the indexes. In term of these combinations,
${\bf R}^{(ax, \,bx)}$ simply reads
\begin{equation}
R^{(ax, \,bx) \b \mu } =
{\textstyle{1 \over 2}}
[ X^{(ax, \sobd bx )  \b   \mu} + (-1)^{n(\b \mu)}
X^{(ax, \sobi bx ) \b \mu^{-1}} ] \label{explr}
\end{equation}
We are now ready to calculate ${\bf R}^{(ax, \,
bx)}(\gamma_o)$. We have
\begin{eqnarray}
&& \hspace{-0.3cm} X^{(ax, \sobd bx ) \b \mu}(\gamma_o) =
\sum_{m=1}^{p-1} \sum_{q=m+1}^{p} \mbox{\large$[$\normalsize}
 T_m^{bx} \,T_q^{ax} \,
X^{\b \mu}([\gamma_{xx}]_1^m \circ \overline{[\gamma_{xx}]}_{m+1}^q
\circ [\gamma_{xx}]_{q+1}^p )  \nonumber \\
&& \hspace{2cm} + (-1)^{n(\b \mu)} T_m^{ax} \,T_q^{bx} \,
X^{\b \mu^{-1}} ([\gamma_{xx}]_1^m \circ \overline{[\gamma_{xx}]}_{m+1}^q
\circ [\gamma_{xx}]_{q+1}^p ) \mbox{\large$]$\normalsize}
\label{xloop}
\end{eqnarray}
where $\overline{[\gamma_{xx}]}_{m+1}^q=\overline{\gamma}_{xx}^{(q)}
\circ \cdots \circ \overline{\gamma}_{xx}^{(m+1)}$ and
$\overline{\gamma}$ indicates the loop $\gamma$ with opposite
orientation. The inversion of the orientation of the
loop (rerouting) in (\ref{xloop})
comes from the following property of the multitangent fields
\begin{equation}
X^{\mu_1 \ldots \mu_n}(\overline{\gamma}) =
(-1)^n X^{\mu_n \ldots \mu_1}(\gamma) \label{invmul}
\end{equation}
This property is a direct consequence of the algebraic constraint
satisfied by the multitangent fields. The reciprocal is not true in
general. Notice that (\ref{invmul}) is the condition imposed
on the elements of the group ${\cal M}_o$.

For the other term in (\ref{explr}) we find
\begin{equation}
(-1)^{n(\b \mu)} X^{(ax, \sobi bx ) \b \mu^{-1}}(\gamma_o) =
X^{(ax, \sobd bx ) \b \mu}(\gamma_o)
\end{equation}
Then
\begin{eqnarray}
&& \hspace{-0.5cm} \psi [{\cal F}_{ab} (x) \times {\bf   R}^{(ax,\,
bx)}(\gamma_o)] =
\int d_{\mu}[A] \, \psi(A) \, Tr[ A_{\b \alpha \b \mu}] \,
{\cal F}_{ab} (x){^{\b \alpha}} \, X^{(ax, \sobd bx ) \b \mu}
(\gamma_o) = \nonumber  \\
&& \hspace{-0.6cm}
2 \! \sum_{m=1}^{p-1} \sum_{q=m+1}^{p} \!\!\! T_m^{[bx,} T_q^{ax]} \!\!
\int \!\! d_{\mu}[A]  \psi(A)  Tr \!\! \left[ \! F_{ab}(x) H_A \!\!
\left\{\!
{\bf R}([\gamma_{xx}]_1^m \circ \overline{[\gamma_{xx}]}_{m+1}^q
\circ [\gamma_{xx}]_{q+1}^p )\! \right\} \!\right] \nonumber  \\
\end{eqnarray}
But
\begin{equation}
Tr[ F_{ab} (x) H_A \{ {\bf R}(\gamma_{xx}) \} ] = \Delta_{ab} (x)
Tr[ H_A \{ {\bf R}(\gamma_{xx}) \} ]
\end{equation}
where $\Delta_{ab} (x)$ is the loop derivative \cite{loopder}. We
conclude
\begin{equation}
\hspace{-0.172cm}
{\cal H}(x) \psi(\gamma_o) = 4 \sum_{m=1}^{p-1} \sum_{q=m+1}^{p}
T_m^{[bx,} T_q^{ax]}  \Delta_{ab}(x)\,
\psi \left( [\gamma_{xx}]_1^m \!\circ \overline{[\gamma_{xx}]}_{m+1}^q
\!\circ [\gamma_{xx}]_{q+1}^p \right)
\end{equation}
This expression corresponds to the usual hamiltonian constraint of
quantum gravity in the loop representation \cite{RG,BP93}. For the
diffeomorphism constraint we obtain a similar result. Equation (\ref{extdif})
reduces to the usual expression of the diffeomorphism constraint in the
loop representation when one particularize this constraint to the case
of loops.

It is important to stress the relationship between the solutions of the
constraints in both representations. Since loops are a
particular case of multitensors, any solution found in the extended
representation can be particularized to loops and would yield in the
limit a solution to the usual constraints of quantum gravity in the loop
representation. The converse is not necessarily true. Given a solution
in the loop representation, it may not generalize to a solution in the
extended representation. An example are the solutions to the
hamiltonian based on smooth nonintersecting loops, which find no
analog in the extended representation.

\section{The formal calculus}
We see that the constraint equations in the extended
representation take a very compact form and they amount, in both
cases, to evaluate the wavefunction on a new object given by the
group product between an element  of the algebra and a combination  of
elements of the group. A point to stress is that these  equations
become also operatives. We can make calculations with  them and the
calculus turns out to be relatively simple. In this section we shall
illustrate this point considering the application of the hamiltonian
constraint over a particular member of a family of solutions in the loop
representation which have a generalization to the extended
representation. We first analyze the nature of these solutions.

In the connection representation based on the Ashtekar variables,
the exponential of the Chern-Simons
form built  with  the Ashtekar connection is a solution of all
the constraints of quantum gravity  with  cosmological constant
\cite{KOD}. This state is given by
\begin{equation}
\Psi_\Lambda^{CS}(A) =
exp (- {\textstyle{12 \over \Lambda}} \!\! \int \tilde{\eta}^{abc} Tr[A_a
\partial_b A_c +{\textstyle{ 2 \over 3}} A_a A_b A_c]\,)
\end{equation}
and we have
\begin{equation}
{\cal H}_\Lambda \, \Psi_\Lambda^{CS}(A) = \left[\, {\cal H} +
{\textstyle{\Lambda \over 6}} det(q)\, \right] \Psi_\Lambda^{CS}(A) = 0
\end{equation}
where $\Lambda$ is the cosmological constant and $q$  the
three-metric. The loop transform of this state is related to  the
expectation value of the Wilson loop \cite{BGPNP}
\begin{eqnarray}
\Psi_\Lambda[\gamma]  &=&  {\textstyle{1 \over 2}} <W(\gamma)> \nonumber \\
&=& exp ( a_{1}[\gamma] \, {\textstyle{\Lambda \over 6}} ) \left[  1 -
({\textstyle{\Lambda \over 6}})^2 \, {\textstyle{3 \over 2}}\, \rho
[\gamma] - ({\textstyle{\Lambda \over 6}})^3  \, 3 \, \tau [\gamma] +
O(\Lambda^{4})   \right]
\label{loopcs}
\end{eqnarray}
The resulting loop wavefunction is the Kauffman Bracket knot polynomial
which is a phase factor times the Jones polynomial. Evaluating the loop
transform (\ref{loopcs}) using perturbative techniques of Chern-Simons
theory \cite{GMM}, explicit expression for the Kauffman Bracket
coefficients can be found. In particular, the phase factor is
proportional to the Gauss self-linking number:
$a_{1}[\gamma]  =  - {\textstyle{3 \over 4}}\varphi [\gamma]$ with
\begin{equation}
\varphi [\gamma] = g_{ax\,by} \, X^{ax}(\gamma) \,X^{bx}(\gamma) \;,
\end{equation}
being
\begin{equation}
g_{ax\,by}   = - {\textstyle{1 \over4\pi}}\epsilon_{abc}
\frac{(x-y)^c}{\mid x-y\mid^3}= - \epsilon_{abc} \, \partial^{\,c}
\nabla^{-2} \, \delta(x-y)
\label{metrica}
\end{equation}
the free propagator of Chern-Simons theory. $\rho[\gamma]$ and
$\tau[\gamma]$ are the
first coefficients of an expansion of the Jones polynomial
in the variable $exp({\textstyle{\Lambda
\over 6}})$ that can be explicitly written as linear functions of the
multitangents with coefficients constructed from $g_{ax\,by}$.

The action of the hamiltonian constraint with cosmological constant can
be evaluated over the Kauffman Bracket polynomial and the analysis of
the resulting equations order by order in $\Lambda$ leads to the
conjecture that the Jones polynomial may be a state of vacuum quantum
gravity \cite{BGP3}. This conjecture was explicitly confirmed for the
first candidate $\rho[\gamma]$ in the loop representation
\cite{BGP2,BGP22} through a laborious (formal) computation. The
expression of this knot invariant in terms of the multitangent fields is
\begin{equation}
\rho [\gamma]  = h_{\mu_1 \mu_2 \mu_3}
X^{\mu_1 \mu_2 \mu_3}(\gamma) +
g_{\mu_1 \mu_3} g_{\mu_2 \mu_4} X^{\mu_1 \mu_2 \mu_3 \mu_4}(\gamma)
\end{equation}
where
\begin{equation}
h_{\mu_1 \mu_2 \mu_3} = \epsilon^{\alpha_1 \alpha_2 \alpha_3} \;
g_{\mu_1 \alpha_1} \, g_{\mu_2 \alpha_2}\, g_{\mu_3 \alpha_3}
\end{equation}
with
\begin{equation}
\epsilon^{\alpha_1 \alpha_2 \alpha_3} = \epsilon^{c_1 c_2 c_3}
\int d^3 t \; \delta(z_1 -t)\,\delta(z_2 -t)\,\delta(z_3 -t)
\end{equation}
The generalization of this knot invariant to extended loops is
straightforward
\begin{equation}
\rho [\gamma] = \rho [{\bf X}(\gamma)]\rightarrow \rho ({\bf X})
\end{equation}
where ${\bf X}$ is now an element of the extended group ${\cal D}_o$.
We now analyze the application of the hamiltonian constraint over this
state in the extended representation. By (\ref{ham}) we have
\begin{eqnarray}
&& \hspace{-1cm} {\cal  H}(x) \,\rho ({\bf R} )  =
2\,h_{\mu_1\mu_2\mu_3}
\left[ {\cal F}_{ab}{^{\mu_1}} (x) \, R^{(ax,\,
bx) \mu_2 \mu_3}
+ {\cal F}_{ab}{^{\mu_1 \mu_2}} (x) \, R^{(ax,\,
bx) \mu_3 } \right]  \nonumber \\
&&
+ 2\,  g_{\mu_1\mu_3}g_{\mu_2\mu_4}
\left[ {\cal F}_{ab}{^{\mu_1}} (x) \, R^{(ax,\,
bx) \mu_2 \mu_3 \mu_4 }
+ {\cal F}_{ab}{^{\mu_1 \mu_2}} (x) \, R^{(ax,\,
bx) \mu_3 \mu_4 } \right]
\end{eqnarray}
We can compute the action of ${\cal F}_{ab}$ over  the  propagators.
The following results are obtained
\begin{eqnarray}
\lefteqn{ \hspace{-0.5cm}
 {\cal F}_{ab}{^{\mu_1 }}(x) \, g_{\mu_1 \mu_3} = -
\epsilon_{ab a_{3}} \delta(x-x_3)
- \partial{_{a_{3}}} g_{ax \, bx_{3}}} \label{R1sobreg} \\
\lefteqn{ \hspace{-0.5cm}
{\cal F}_{ab}{^{\mu_1 \mu_2}}(x) \, g_{ \mu_1 \mu_3 }
g_{\mu_2 \mu_4}  =   g_{\mu_3  [ ax } \, g_{\, bx] \, \mu_4 }} \\
\lefteqn{ \hspace{-0.5cm}
{\cal F}_{ab}{^{\mu_1}}(x) \, h_{\mu_1 \mu_2 \mu_3 } =
- g_{\mu_2  [\,ax} g_{\, bx] \, \mu_3} + (g_{ax \, bx_{2}} -
g_{ax \, bx_{3}}) g_{\mu_2 \mu_3}}  \nonumber\\
&& \hspace{2.45cm}
+ {\textstyle{1 \over 2}} g_{ax  \,  bz}  \epsilon^{def}  [g_{\mu_3 \,  dz}
\partial_{a_{2}}  \, g_{ex_{2} \, fz} - g_{\mu_2 \,  dz}
\partial_{a_{3}}  \, g_{ex_{3} \, fz} ] \label{uno}\\
\lefteqn{ \hspace{-0.5cm}
{\cal F}_{ab}{^{\mu_1 \mu_2}}(x) \, h_{ \mu_1 \mu_2 \mu_3}
 =   2\, h_{ax \,  bx \, \mu_3}}
\end{eqnarray}
In the last term of equation (\ref{uno}) an  integral
in  $z$  is  assumed.  The  derivatives  that  appear   in    the
above expressions can  be applied over  the  ${\bf R}$'s
integrating by parts, and using the differential  constraint  we
generate from them terms of lower rank. For example from
(\ref{R1sobreg}) we have
\begin{equation} g_{\mu_2\mu_4}
\partial_{a_{3}} g_{ax \, bx_{3}} R^{(ax, \,bx) \mu_2 \mu_3
\mu_4 } = g_{\mu_2\mu_4} (g_{ax \, bx_{2}} -g_{ax \, bx_{4}}) R^{(ax,
bx) \mu_2  \mu_4 }
\end{equation}
Performing these
calculations, the following partial  results  are obtained for each of
the four expressions quoted above:
\begin{eqnarray*} \lefteqn{
\hspace{-0.7cm} 1)\,  - \epsilon_{abc}  g_{\mu_1 \mu_2} R^{(ax,
bx) \mu_1 \,cx\, \mu_2 } - (g_{ax \, bx_{1}} -g_{ax \,
bx_{2}}) g_{\mu_1\mu_2} R^{(ax, \,bx) \mu_1  \mu_2 }}  \\
\lefteqn{ \hspace{-0.7cm} 2)\;\;\, g_{\mu_1  [ ax } \, g_{\, bx] \,
\mu_2 } R^{(ax, \,bx) \mu_1 \mu_2}} \\ \lefteqn{
\hspace{-0.7cm} 3)\, -  g_{\mu_1  [\,ax} g_{\, bx] \, \mu_2} R^{(ax,
bx) \mu_1 \mu_2 } +   (g_{ax \, bx_{1}} -g_{ax \, bx_{2}})
g_{\mu_1 \mu_2} R^{(ax, \,bx) \mu_1 \mu_2 }} \\ &&
\hspace{4cm} - \epsilon^{def} g_{ax  \,  bz}   g_{\mu_1 \,  dz}  g_{ex
\, fz} R^{(ax, \,bx) \mu_1 } \\ \lefteqn{ \hspace{-0.7cm}
4)\;\;\,    2\, h_{ax \,  bx \, \mu_1} R^{(ax, \,bx) \mu_1 } }
\end{eqnarray*}
Some contributions cancel each other and we finally obtain
\begin{eqnarray} {\cal  H}(x) \,\rho
({\bf R} )  &=& - 2\; \epsilon_{abc}  g_{\mu_1 \mu_2} R^{(ax,
bx) \mu_1 \,cx\, \mu_2 } \nonumber \\ && + 2\; [ 2\, h_{ax \,
bx \, \mu_1} - \epsilon^{def} g_{ax  \,  bz}   g_{\mu_1 \,  dz}  g_{ex
\, fz}] R^{(ax, \,bx) \mu_1 }  \label{hamrho}
\end{eqnarray}
In the above expression, the square bracket vanishes identically because
the  second term is nothing else that other form to  express the  first
one. One can easily check this fact. Developing
$R^{(ax, \,bx) \mu_1 \,cx\, \mu_2}$ we get
\begin{equation}
R^{(ax, \,bx) \mu_1 \,cx\, \mu_2} =  -2 \, R^{(ax\,bx\,\mu_1
\,cx\, \mu_{2})_c} + R^{(cx\,ax\,\mu_1 \,bx\, \mu_{2})_c} +
R^{(bx\,cx\,\mu_1 \,ax\, \mu_{2})_c} \label{Rrango5}
\end{equation}
and the contribution of the rank five term vanishes due to symmetry
considerations. We conclude
\begin{equation}
{\cal  H}(x) \,\rho ({\bf R} )= 0
\end{equation}
We see that the explicit computation of this formal
result in the extended representation involves only a few simple
steps. The power of the method is reflected also in the fact that one
can put forward the analytical computation of the hamiltonian constraint
over the next candidate to a nondegenerate solution suggested by
Br\"ugmann, Gambini and Pullin. The expression of the third coefficient
of the Jones polynomial in terms of the elements of the extended group
is \cite{DG}
\begin{eqnarray}
\tau ({\bf X}) &=& [h_{\mu_1\mu_2\alpha}
g^{\alpha\beta} h_{\mu_3\mu_4\beta} - h_{\mu_1\mu_4\alpha}
g^{\alpha\beta} h_{\mu_2\mu_3\beta}] \, X^{\mu_1\mu_2\mu_3\mu_4} +
\nonumber \\ & & \hspace{2cm} g_{(\mu_1\mu_3}h_{\mu_2\mu_4\mu_5)_c} \,
X^{\mu_1\mu_2\mu_3\mu_4\mu_5}+ \nonumber \\ & &
[2g_{\mu_1\mu_4}g_{\mu_2\mu_5}g_{\mu_3\mu_6} + {\textstyle{1 \over 2}}
g_{(\mu_1\mu_3}g_{\mu_2\mu_5}g_{\mu_4\mu_6)_c}] \,
X^{\mu_1\mu_2\mu_3\mu_4\mu_5\mu_6}
\end{eqnarray}

\vspace{0.3cm}
\noindent
In the extended loop representation we have been able to prove that
\begin{equation}
{\cal  H}(x) \,\tau ({\bf R} ) = 0
\end{equation}
explicitly \cite{ultima}. This is a new result coming from the
formulation of Quantum Gravity in the extended loop space.

\section{The issue of the regularization}
The extended representation provides a new scenario to analyze the
regularization problem in quantum gravity. In the loop representation
regularization ambiguities appear both at the level of quantum operators
{\it and} of quantum states. Whereas the first problem is common to all
the representations that one can construct for quantum gravity (and lies
in the fact that the constraints involve the product of operators
evaluated at the same point), the second is typical of the loop
representation. In the case of quantum gravity the loop wavefunctions
are knot invariants and their analytic expressions require the
introduction of a regularization (framing). This difficulty not only
arises for the gravitational case, even in the simple case of a free
Maxwell field \cite{DGNT}, it is known that the quantum states in the
loop representation are ill defined and a regularization is needed.

In the extended representation the second difficulty can be solved. We
are going to show that with an adequate restriction of the domain of
dependence, the extended wavefunctions are smooth functionals of the
variables. In what concerns the regularization of the constraints, we
shall limitate the analysis to the case of wavefunctions with a totally
specified analytical dependence. More precisely, we shall study the
action of the regularized hamiltonian constraint over the wavefunctions
that are formally annihilated by the constraint. A more general
discussion of the regularization and renormalization of the constraints
as well as the consistency of their algebra will be given elsewhere
\cite{AC}.

\subsection{The smoothness of the extended wavefunctions}
In the loop representation the coefficients of the expansion
(\ref{loopcs}) of the expectation value of the Wilson loop in terms of
the cosmological constant are knot invariants. It is easy to see that
their generalization to the extended representation are also
diffeomorphism invariants. For that, consider the extended loop
transform of the exponential of the Chern-Simons form
\begin{equation}
\Psi_{\Lambda}({\bf X})=\int d_{\mu}[A] \, {\em e}^{S_{\Lambda}(A)} \,
Tr[{\bf A} \cdot{\bf
X}] = \sum_{n=0}^{\infty} [{\bf g}^{(n)} \cdot {\bf X}]\;\Lambda^n
\label{extdiftransf}
\end{equation}
where the dot indicates the contraction of indexes. As the measure of
integration  and  the  Chern-Simons  action  are diffeomorphism
invariants, we have
\begin{equation}
\Psi_{\Lambda}({\bf X})=\int d_{\mu}[A_D] \, {\em e}^{S_{\Lambda}(A_D)} \,
Tr[{\bf A}_D \cdot \Lambda_D \cdot {\bf X}]
\label{extdiftransf2}
\end{equation}
where $\Lambda_D \cdot {\bf X}$ and ${\bf A}_D={\bf A}_o \cdot
\Lambda_{D^{-1}}$ are the transformed quantities under the diffeomorphism
$x'^a = D^a (x)$. ${\bf A}_D$ is in another gauge. Due to the fact that
the measure of integration, the Chern-Simons action and the trace of the
extended holonomy are invariant under gauge transformations connected
with the identity, we conclude $\Psi_{\Lambda}({\bf
X})=\Psi_{\Lambda}(\Lambda_D \cdot {\bf X})$. From (\ref{extdiftransf})
we get for any $n$
\begin{equation}
{\bf g}^{(n)} \cdot {\bf X}  =
{\bf g}^{(n)} \cdot \Lambda_D \cdot {\bf X}
\label{staten}
\end{equation}
The coefficients are then invariant under diffeomorphism
transformations. Note that this result does not imply that
$\rho({\bf X})$ or $\tau({\bf X})$ are diffeomorphism invariants.
{}From (\ref{loopcs}) the state (\ref{staten}) contains to any order
in the cosmological constant contributions of different knot
invariants. The invariance of these coefficients has to be checked
explicitly.

Let us consider now the regularity properties of the extended
wavefunctions. Generically the multitensors $X^{\b \mu}$ are
distributional, as it is directly inferred from (\ref{diffconst}).
Any multitensor that satisfies the differential constraint can be
written in the form ${\bf X} = \SG[\phi] \cdot {\bf Y}$, where the ${\bf
Y}$ fields satisfy the homogeneous differential constraint.
For example, for the rank two component we have
\begin{equation}
X^{ax\,by} = Y^{ax\,by} + \phi_{ \phantom{Ai} y}^{\;ax} \; Y^{by} -
\phi_{ \phantom{Ai} x}^{\;by} \; Y^{ax} -
\phi_{ \phantom{Ai} z}^{\;ax} \phi_{ \phantom{Ai}
z,c}^{\;by} \, Y^{cz} + \phi_{ \phantom{Ai} o}^{\;[by} \; Y^{ax]}
\label{rank2}
\end{equation}
The function $\phi$ fixes a prescription for the decomposition of the
multitensors in transverse and longitudinal parts: ${\bf Y}=
\delta_T \cdot {\bf X}$ with
\begin{eqnarray}
\delta\ind{T}{\vc\mu 1n}{\vc\nu 1m}  &=&
\delta_{n,m} \;\;
\delta\ind{T}{\mu_1}{\nu_1}\cdots \delta\ind{T}{\mu_n}{\nu_n} \\
\delta_{T \phantom{A} by}^{\phantom{A} ax}  &=&
\delta_{\phantom{AA} by}^{ax} - \phi_{ \phantom{Ai} y,b}^{\;ax}
\end{eqnarray}
As the ${\bf Y}$'s satisfy the homogeneous differential constraint,
they can be assumed to be smooth functions. In this case, all the
divergent behavior of the ${\bf X}$ is concentrated in the function
$\phi$. The $\SG$'s control then the divergent character of the
group elements.

Let us define  the following set of elements of the extended space:
${\bf X} \in \{{\bf X}\}_s$ if, and only if, there exists a
prescription function $\phi$ such that $\delta_{T}[\phi]
\cdot{\bf X}= {\bf Y}$ is a smooth function. We shall demonstrate
that the wavefunctions defined on this domain are smooth in the
extended variables and that this property is invariant under
diffeomorphism transformations.

Given a diffeomorphism transformation $\Lambda_D$, it can be shown
that $\delta_{DT}\equiv \Lambda_{D^{-1}} \cdot \delta_{T} \cdot
\Lambda_D$ is a transverse projector in the prescription
\begin{equation}
\phi_{D \phantom{Ai} y}^{\phantom{A} ax}  =  J(x)  \; \frac
{\partial x^a}
{\partial D^b (x)} \phi_{\phantom{AAA} D(y)}^{\phantom{} bD(x)}
\end{equation}
where $J(x)$ is the jacobian of the coordinate transformation and $\phi$
the function that fixes the prescription of the projector $\delta_T$
\cite{GGG}. In this prescription ${\bf X}=\SG \cdot {\bf Y}=
\Lambda_{D^{-1}} \cdot \SG_{D^{-1}} \cdot \Lambda_D \cdot {\bf Y}$.  For
any diffeomorphism transformation $\Lambda_{D}$, the transverse part of
$\Lambda_{D} \cdot {\bf X}$ is a smooth function with the prescription
$\phi_{D^{-1}}$. In effect
\begin{equation}
\delta_{D^{-1}T} \cdot ( \Lambda_{D} \cdot {\bf X} ) =
\delta_{D^{-1}T} \cdot \SG_{D^{-1}} \cdot \Lambda_D \cdot {\bf Y} =
\Lambda_{D} \cdot {\bf Y}
\end{equation}
The set $\{{\bf X}\}_s$ is then invariant under diffeomorphism
transformations. For any ${\bf X} \in \{{\bf X}\}_s$ we have
\begin{equation}
\psi({\bf X}) ={\bf g} \cdot {\bf X} ={\bf g} \cdot \SG [\phi]
\cdot {\bf Y} \equiv {\bf g}_{\phi} \cdot {\bf Y}
\end{equation}
All the distributional character of the wavefunction is concentrated in
the ``metric" ${\bf g}_{\phi}$. They are well defined functionals,
smooth in the extended variables. Moreover, the diffeomorphism
transformed of the wavefunction would be defined on the same domain
and is also smooth.

\subsection{The regularization of the constraints}

We shall restrict the analysis of the regularization to the case of
the point splitting method, based on a delocalization of the point
where the operators are evaluated. The point splitting version of
the constraints are
\begin{eqnarray}
&&{\cal C}^{\;\epsilon}_{ax}\,\psi ({\bf R})=
\int \!d^3 w \!\!\int \!d^3 v
\,f_{\epsilon} (w,x)\, f_{\epsilon} (v,x)\,
\psi({\cal  F}_{ab}(w) \times {\bf R}^{(bv)} )  \label{regdif}\\
&&{\cal H}^{\,\epsilon} (x) \, \psi({\bf R}) = \nonumber \\
&&2\int \!d^3 w \!\!\int d^3 u \!\!\int \!d^3 v \,f_{\epsilon} (w,x)
\,f_{\epsilon} (u,x) \,f_{\epsilon} (v,x)\,
\psi ({\cal F}_{ab} (w) \times {\bf   R}^{(au, \,bv)})
\end{eqnarray}
where $f_{\epsilon}$ is any appropriate symmetric smearing of the delta
function. Notice that this point splitting regularization is not
uniquely determined by the formal factor ordered expression. Several
sources of ambiguities arise, the first one is related with the
background metric used in the smearing functions. It is also possible,
but not necessary, to preserve the gauge invariance in the
regularization process. Finally additional factor ordering problems may
arise due to the distributional character of the fundamental fields.
A more complete and lengthier discussion will be given elsewhere
\cite{AC}. Here we shall proceed as follows: we shall introduce a naive
point splitting and study the action of the regularized and renormalized
operators on the formal solutions. We shall prove that there is a factor
ordering that insures the consistency between the known results in the
connection and the loop representation.

In the previous section we have shown that the invariance under
diffeomorphisms of the coefficients of the expansion of the generalized
transform (\ref{extdiftransf}) is insured by construction. Also that,
with an appropriate definition of the domain of dependence, the
wavefunctions can be endowed with convenient regularity properties (in
particular, the smoothness dependence on the extended variables can be
insured in a diffeomorphism invariant way). We check now the good
behavior of the regularized diffeomorphism constraint for the particular
case of the Gauss invariant $\varphi ({\bf R})$. From (\ref{regdif}) we
obtain
\begin{equation}
{\cal C}^{\;\epsilon}_{ax}\,\varphi ({\bf R})
= \int \!d^3 w \!\!\int \!d^3 v
\,f_{\epsilon} (w,x)\, f_{\epsilon} (v,x)\, g_{\mu_1 \mu_2}\, {\cal
F}_{ab}{^{\mu_1}}(w) \, R^{(bv) \mu_2}
\end{equation}
This result is valid for any prescription. Due to practical
computational reasons we shall restrict the domain of the
wavefunctions to those prescriptions connected by diffeomorphisms to
the ``transverse" prescription, given by
\begin{equation}
\phi_{o \phantom{Ai} y}^{\phantom{A} ax}  =  {\frac {1}{4\pi}} \; \frac
{\partial}
{\partial x_a} \frac {1}{\mid x-y \mid}
\end{equation}
In the transverse prescription the free Chern-Simons propagator $g_{ax\,by}$
takes the form (\ref{metrica}). Then using (\ref{R1sobreg}) we get
\begin{equation}
{\cal
C}^{\;\epsilon}_{ax}\,\varphi ({\bf R})=- \epsilon_{abc} \int \!d^3 w
\!\!\int \!d^3 v \,f_{\epsilon} (w,x) \,f_{\epsilon} (v,x)\, R^{(bv)\,
cw} \label{regdifgauss}
\end{equation}
where
\begin{equation}
R^{(bv)\, cw} =  Y^{bv\,cw} + Y^{cw\,bv}
\end{equation}
is a smooth function symmetric under the interchange of
the indexes $b$ and $c$ (using the fact that the integration  points are
undistinguishable). The last expression is well defined and we conclude
that
\begin{equation}
{\cal  C}^{\;\epsilon}_{ax}\,\varphi ({\bf R})=0
\end{equation}
Notice that no divergences occur in (\ref{regdifgauss}) and we
do not need to take the limit when $\epsilon$ goes to zero. The
diffeomorphism constraint is perfectly well defined and no
renormalization is needed. A similar result holds for the other known
invariants.

Let us analyze now the action of the regularized hamiltonian
constraint on the Alexander-Conway coefficient $\rho({\bf
R})$. We get in this case
\begin{eqnarray}
&& \hspace{-1.9cm}{\cal  H}^{\,\epsilon} (x) \,\rho ({\bf R} )  =
\int \!d^3 w \!\!\int d^3 u \!\!\int \!d^3 v \,f_{\epsilon} (w,x)
\,f_{\epsilon} (u,x) \,f_{\epsilon} (v,x)\nonumber \\
&&\hspace{-0.8cm}\{- \; \epsilon_{abc}  g_{\mu_1 \mu_2}
R^{(au, \,bv) \mu_1 \,cw\, \mu_2 } \,+\nonumber \\
&& \hspace{0.3cm} [ 2\, h_{aw \,  bw \, \mu_1}
- \epsilon^{def} g_{aw  \,  bz}   g_{\mu_1 \,  dz}  g_{eu \, fz}]
R^{(au, \,bv) \mu_1 } \,+ \nonumber \\
&&\hspace{4cm}
(g_{aw\,bu}-g_{aw\,bv})\,g_{\mu_1 \mu_2} R^{(au\,\mu_1\,bv\,\mu_{2})_c}\}
\label{hamregrho}
\end{eqnarray}
The last term in the above expression (what we call the ``anomalous
term") appears as a consequence of the modification of the differential
constraint of ${\bf R}^{(au,\,bv)}$. We have now
\begin{eqnarray}
&&\hspace{-0.8cm}
 \partial_{\mu_i} R^{(au, \,bv) \mu_1 \ldots \mu_i
\ldots \mu_n } =
\left[ \delta(x_i-x_{i-1}) - \delta(x_i-x_{i+1}) \right]
R^{(au, \,bv) \mu_1 \ldots \ra \mu_i \ldots \mu_n } \nonumber \\
&&\hspace{0.25cm} + \left[ \delta(x_i-u) - \delta(x_i-v) \right]
(-1)^{n-i} R^{(au\,\mu_1 \ldots \mu_{i-1}\,bv\,\mu_n \ldots
\mu_{i+1})_c}
\end{eqnarray}
instead of the normal differential constraint. In the above expression,
$x_0=u$ and $x_{n+1}=v$. The calculation of the regulated terms involves
the consideration of the divergences that comes from the group elements
(through the matrix sigma) and from the metric terms. The first
observation is that both types of contributions are of the same order.
Let us consider for example the divergent contributions associated with
the rank five group elements contained in $R^{(au, \,bv)\mu_1 \,cw\,
\mu_2}$. One can see that all these terms are equivalent in their
divergent character to
\begin{equation}
\phi_{o \phantom{Ai} v}^{\,\;au}\,Y^{(bv \,\mu_1 \,cw\, \mu_{2})_c}
\end{equation}
with $Y^{(bv \,\mu_1 \,cw\, \mu_{2})_c}$ a regular function in the limit
$\epsilon \rightarrow 0$. Notice that this expression gives the leading
divergence of the rank five term in (\ref{hamregrho}). But
\begin{equation}
\epsilon_{bca} \phi_{o \phantom{Ai} v}^{\,\;au}\,Y^{(bv \,\mu_1 \,cw\,
\mu_{2})_c}= g_{bu\,cv}\,Y^{(bv \,\mu_1 \,cw\,\mu_{2})_c}
\end{equation}
is exactly the same contribution that comes from the anomalous term.
Notice that in this case $g_{\mu_1 \mu_2}\,R^{(bv \,\mu_1 \,cw\,
\mu_{2})_c}$ is a regular function in the limit $\epsilon \rightarrow 0$
due to the point splitting does not affect two consecutive indexes.

The result (\ref{Rrango5}) ensures that the contribution of the rank five
term in the hamiltonian vanishes due to the same symmetry properties
used in the formal calculus. One can see that the second
term in (\ref{hamregrho}) also vanishes when one removes the
regulators. For the anomalous term we get
\begin{eqnarray}
&&2\int \!d^3 w \!\!\int d^3 u \!\!\int \!d^3 v \,f_{\epsilon} (w,x)
\,f_{\epsilon} (u,x) \,f_{\epsilon} (v,x)\,
g_{aw\,bv}\,g_{\mu_1 \mu_2} R^{(au\,\mu_1\,bv\,\mu_{2})_c}=\nonumber\\
&&\hspace{2cm}\frac{2}{\sqrt{2\pi}\,\epsilon} \,\epsilon_{abc} \,
g_{\mu_1 \mu_2}\;\partial^{cy}\,
R^{(ax\,\mu_1\,by\,\mu_{2})_c} \mbox{\large{$\mid$}}_{y=x}
+ O(\epsilon)
\end{eqnarray}
where we have used a gaussian regulator $f_{\epsilon}
(\vec{z})=(\sqrt{\pi} \epsilon )^{-3}\, exp\,(- z^2 \epsilon^{-2})$. Then
\begin{eqnarray}
{\cal  H}_{r} (x) \,\rho ({\bf R} )  &=&
\lim_{\epsilon \rightarrow 0} \;\epsilon \,
{\cal  H}^{\, \epsilon} (x) \,\rho ({\bf R} ) \nonumber  \\
&=& {\textstyle{\sqrt{\frac{2}{\pi}}}} \;\epsilon_{abc} \; g_{\mu_1 \mu_2}\;
\partial^{cy}\,
R^{(ax\,\mu_1\,by\,\mu_{2})_c} \mbox{\large{$\mid$}}_{y=x}
\end{eqnarray}
We conclude that the renormalized hamiltonian constraint does not
annihilate the generalized diffeomorphism invariant corresponding to the
second coefficient of the Alexander-Conway knot polynomial in the
(naive) point splitting regularization procedure.

A compatibility argument arises from the beginning. We know that the
exponential of the Chern-Simons form is an {\it exact} quantum state of
the hamiltonian constraint with cosmological constant in the connection
representation. What about the regulated equations order by order in
$\Lambda$ that correspond to the same state in the extended
representation? As it was just mentioned, the formal calculus works in
the expected way. But the point slitting regularization makes these
equations to fail. However, the consistency can be restored by
introducing a counterterm in the hamiltonian.

A counterterm is a regularized term which has not effect over a regular
Wilson functional (that is to say, over an extended Wilson functional
constructed with non distributional connections). Consider for example
the following expression, symmetric under the interchange of
the internal indexes
\begin{eqnarray}
&&\hspace{-1.5cm}(A^i_{aw} A^j_{bu} - A^i_{aw} A^j_{bv})
\frac{\delta}{\delta  A^{(j}_{bv}} \frac{\delta}{\delta A^{i)}_{au}}
W_A ({\bf R}) =\nonumber \\
&&\hspace{0.8cm}
\{\,Tr[A_{(aw\,\mid \, \b \mu \mid \,bu) \, \b \nu}]-Tr[A_{(aw\, \mid \,
\b \mu \mid \, bv) \,\b \nu}]\,\}
\,R^{(au\, \b \mu \,bv \, \b \nu)_c}
\end{eqnarray}
It is clear that this term vanishes in the limit $\epsilon \rightarrow
0$ if the connections are regular functions, but it may have a
nontrivial contribution if the connections are distributions. The
corresponding regularized expression in the extended space is
\begin{equation}
{\em C}^{\,\epsilon} = R^{(au \,\b \mu \,bv \,\b \nu)_c}
(\frac{\delta}{\delta R^{(aw \,\mid\, \b \mu \mid\, bu) \,\b \nu}} -
\frac{\delta}{\delta R^{(aw \,\mid \, \b \mu \mid \,bv)\, \b \nu}})
\end{equation}
This term generates anomalous type contributions. For example
\begin{equation}
{\em C}^{\,\epsilon} (g_{\mu_1 \mu_2} \, R^{\mu_1 \mu_2})=
2 (g_{aw\,bu}-g_{aw\,bv})\,R^{(au \,bv)_c}
\end{equation}
The addition of the counterterm ${\em C}^{\,\epsilon}$ in ${\cal H}$
insures that the Kauffman Bracket polynomial is annihilated (up to the
second order in $\Lambda$) by the renormalized hamiltonian constraint
with cosmological constant. This is an important result. In the extended
representation one can prove that the Kauffman Bracket polynomial is a
quantum state of gravity with cosmological constant including
regularization. Notice that the introduction of the counterterm ${\em
C}^{\,\epsilon}$ is equivalent to a particular choice of ordering of the
connections in the hamiltonian constraint.

\section{Conclusions}
An extended loop representation for quantum gravity was constructed. The
formal calculus associated with the constraints have been developed and
the regularization problems discussed.

We have shown that the extended loops provide an appropriate arena for
the definition of the quantum states and that the constraint equations
present practical calculational advantages. These facts can be
considered as an improvement of the loop representation in the
description of quantum gravity. However in this approach one of the most
attractive ingredients of the loop representation has been lost: the
fact that the diffeomorphism constraint was easily solvable in terms of
knots. A further study of the characterization of diffeomorphism
invariant classes of extended loops is in order. In this sense, the fact
that the knot  polynomial  solutions  have  direct analogs in the new
representation can be viewed, in our opinion, as a suggestion that the
topological and geometric insights of the loop representation be
inherited in some sense by their extension to the generalized space.

One may wonder about the equivalence between all the possible
representations that can be formulated in the extended space. As we just
mentioned, representations associated with the groups ${\cal M}_o$ and
${\cal X}_o$ can be developed in a way similar to the ${\cal D}_o$ case.
Moreover, it is still unclear which among all possible representations
behave as the dual of the connection representation. These
representations obey different types of first class constraints. In the
case of ${\cal D}_o$ these constraints are the linearity constraint and
the diffeomorphism and hamiltonian constraints. The presence of other
constraints besides the usual gravity constraints allow to eliminate the
superfluous degrees of freedom of the theory. One can see that the
elimination is complete in ${\cal X}_o$, where all the gauge ambiguity
disappears and, in consequence, all the representations can be
considered as essentially equivalent.

An important property of the extended representation of quantum General
Relativity (and in general of any gauge theory) is that it provides, in
a natural way, a framework to develop a classical counterpart of the
quantum theory. One can view the role of the multitensors as
configuration variables of a canonical theory. The conjugate momenta are
represented by functional derivatives. This suggests that there exists
an underlying classical  Hamiltonian theory that under canonical
quantization yields directly the extended loop representation. This was
unclear with loops, where the loop representation could only be
introduced through a noncanonical quantization. For the Maxwell case
this theory was studied \cite{classical} and found to be equivalent to
the usual Maxwell theory. For the nonabelian case it is yet to be studied.

In what concerns the regularization procedure some comments are in
order. In first place it is important to point out that a gauge
invariant regularization procedure may be introduced. However this
regularization is totally equivalent to the naive regularization we
have considered here. Secondly, it is important to remark that the
generalized knot invariants are well defined in the space of ${\bf
R}$'s that are related with continuous ${\bf Y}$'s for some
prescription ${\phi}$ and therefore in a diffeomorphism invariant
background independent domain. However we need to impose, for
practical reasons, a further, background dependent, restriction in
the domain of the wavefunctions: the transverse prescription. This
prescription allows to perform the explicit analysis of the
regularization and renormalization of the constraints. Therefore
the regularization procedure is clearly background dependent; the
smearing functions, the Chern-Simons propagators and the extended
space are in the transverse prescription.

A more geometric definition of the relevant extended space, for
instance by restricting the domain of the wavefunctions to some
geometric smearing of loops (bands or tubes), could allow to keep
for one side well defined knot invariants and on the other side to
analyze the action of the regularized constraints without any
reference to a background metric.

\section*{Acknowledgments}
We wish especially to thank Jorge Pullin for discussions of some of the
physical implications of the extended representation. We also thank
Daniel Armand Ugon for a critical reading of the manuscript. R. G.
wishes to thank Abhay Ashtekar, Jose Mour$\tilde{\mbox{a}}$o, Carlo
Rovelli, Lee Smolin and Madhavan Varadarajan for fruitful discussions.



\begin{thebibliography}{99}
\bibitem{reruteo} Reroutings are the main effect of the nonlinear terms
in Yang-Mills theory and Quantum Gravity. See references \cite{GT,RS}.
\end{thebibliography}
\end{document}